%% file: main.tex
\PassOptionsToPackage{table}{xcolor}
\documentclass[sigconf, nonacm]{acmart}
\newcommand\vldbdoi{XX.XX/XXX.XX}

\newcommand\vldbavailabilityurl{URL_TO_YOUR_ARTIFACTS}
 
\usepackage{multirow}
\usepackage{enumitem}
\usepackage{adjustbox}
\setlist[itemize]{leftmargin=*}
\newcommand{\Paragraph} [1] {\smallskip\noindent{\bf #1. }}

\begin{document}
\title{Towards Autonomous Graph Data Analytics with Analytics-Augmented Generation}


\author{Qiange Wang, Chaoyi Chen, Jingqi Gao, Zihan Wang,  Yanfeng Zhang, Ge Yu}\thanks{Qiange and Chaoyi contributed equally. Yanfeng is the corresponding author}
\affiliation{
  \institution{Northeastern University\country{China} 
  }
}
\email{{wangqg,zhangyf,yuge}@mail.neu.edu.cn;{chenchaoy,gaojingqi}@stumail.neu.edu.cn}

\begin{abstract}
This paper argues that reliable end-to-end graph data analytics cannot be achieved by retrieval- or code-generation–centric LLM agents alone. Although large language models (LLMs) provide strong reasoning capabilities, practical graph analytics for non-expert users requires explicit analytical grounding to support intent-to-execution translation, task-aware graph construction, and reliable execution across diverse graph algorithms. We envision Analytics-Augmented Generation (AAG) as a new paradigm that treats analytical computation as a first-class concern and positions LLMs as knowledge-grounded analytical coordinators. By integrating knowledge-driven task planning, algorithm-centric LLM–analytics interaction, and task-aware graph construction, AAG enables end-to-end graph analytics pipelines that translate natural-language user intent into automated execution and interpretable results. 


\end{abstract}

\maketitle






\input{sec1.tex}
\input{sec2.tex}

\input{sec3.tex}

\input{sec4.tex}

\input{sec5.tex}

\bibliographystyle{ACM-Reference-Format}
\bibliography{sample}

\end{document}

%% file: sec1.tex
\begin{figure}
  \centering
  \includegraphics[width=1.05\linewidth]{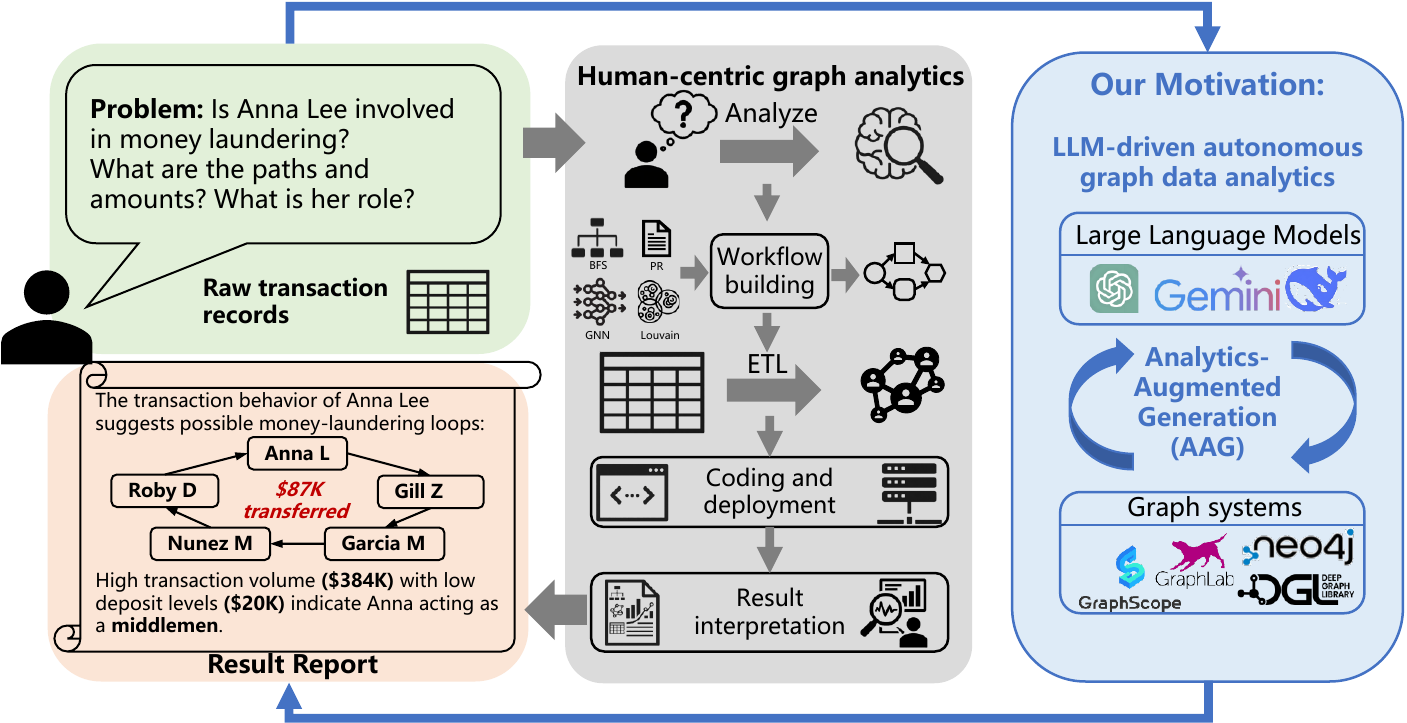}
  \vspace{-0.15in}
  \caption{An example of human-centric graph data analytics and the motivation of our design.
  }
  \label{fig:sec1_problem}
  \vspace{-0.2in}
\end{figure}

\section{Introduction}

Research on graph data analysis has spanned decades, yielding a rich and diverse ecosystem of analytics, mining, and learning algorithms~\cite{gnn_survey_arxiv2019,GraphAnalysisSurvey_CSUR_2025,GSM_SIGMOD_2024}. However, the semantic complexity of graph-structured data, together with the breadth of available graph algorithms, has created a substantial gap between massive graph data and effective real-world usage. 

Figure \ref{fig:sec1_problem} illustrates a human-centric graph analytics workflow for money laundering ctivity detection over large-scale transaction data. Starting from an investigative question, analysts must manually interpret the analysis objective, decompose it into a multi-step analytical workflow, select appropriate graph algorithms for each stage, and construct task-specific input graphs from heterogeneous raw records. Finally, analysts implement, deploy, and iteratively tune the analysis program to to obtain results, which are ultimately consolidated into a human-readable analytical report.
This end-to-end process remains heavily dependent on specialized expertise in graph data management, analytics, and application, resulting in a high barrier for general practitioners~\cite{LLM4Graph_NIPS_2024}. Consequently, despite their strong expressive power for modeling deep and complex relationships, graph-based analytical techniques remain far less adopted in practice than relational analytics \cite{Andy_report_2024}, leaving a large fraction of the value embedded in large-scale, highly connected data underexplored in real-world applications.

Recent advances in data agents combine high-level LLM-based coordination with low-level deterministic analytical data processing \cite{liu2025joyagent,hong2025data,guo2024ds,zhang2025deepanalyzeagenticlargelanguage,sun2025agenticdata,DataAgentSurvey_ARXIV_2025,LLMasDataAnalystSurvey_ARXIV_2025}. While effective for structured workloads, this paradigm does not readily extend to graph analytics, where analytical workflows, data representations, and algorithmic behaviors are inherently more complex. First, existing works require analytical tasks to be specified in imperative, algorithm-specific forms (e.g., searching for high-traffic cyclic structures) rather than as high-level intents (e.g., identifying money-laundering behaviors). Consequently, intent-to-instruction translation remains a manual responsibility of human experts. Second, graph data do not naturally exist in an analysis-ready form and must be constructed from raw data, considering task-aware schema extraction (e.g., selecting bank transaction records while excluding routine payments for fraud detection) and structural organization (e.g., property graph or CSR), which demands substantial manual effort and is often overlooked in existing work \cite{chat2graph,GraphTeam_arxiv_2024,GraphAgentReasoner_arxiv_2024}. Third, at the execution level, the inherent complexity and diversity of graph algorithms limit the effectiveness of code-generation-based analytics mechanisms built around a single model or narrow APIs (e.g., GQL \cite{GQL_SIGMOD_2022} or GAS models\cite{GraphAgentReasoner_arxiv_2024, PowerGraph_OSDI_2012}), which typically support only restricted classes of algorithms and may yield fragile or unreliable results. These issues become even more pronounced in complex, multi-stage analytical workflows, where intermediate results, execution plans, and data dependencies must be explicitly managed across stages.

In this work, we envision \textbf{A}nalytics-\textbf{A}ugmented \textbf{G}eneration (\textbf{AAG}) as a new paradigm for constructing autonomous graph data analytics agents, as illustrated in Figure~\ref{fig:sec1_problem}. AAG treats analytical computation as a first-class concern and position LLM as the intelligent coordinator, which is responsible for interpreting user intents, synthesizing analytical workflows, guiding data extraction and preparation, orchestrating execution across analytical modules, and explaining analytical results. To effectively support workflow construction, data preparation, and LLM–analytics interaction, AAG is built around three key modules.

\begin{itemize}
\item \textbf{Knowledge-driven task planning.} Rather than relying solely on the model’s internal knowledge, AAG integrates LLM reasoning with retrieval-augmented generation over a purpose-built, structured knowledge base to translate high-level user intents into concrete planning decisions, including analytical workflow construction, algorithm selection, and parameter configuration.

\item \textbf{Algorithm-centric LLM--analytics interaction.}  Rather than relying on unconstrained code generation  in fixed languages,
AAG adopts an algorithm-centric interaction model that allows LLMs to invoke and compose verifiable graph analytics modules in graph systems. This design enables flexible orchestration while ensuring reliable and reproducible execution.

\item \textbf{Task-aware graph construction.}
Instead of indiscriminately extracting graphs from raw data sources, AAG derives task-aware schemas to selectively construct analysis-relevant graphs, avoiding irrelevant structures that may interfere with analysis. The resulting graphs are further organized into execution-friendly representations aligned with algorithmic requirements, enabling efficient and dependable graph analytics.
\end{itemize}

The remainder of this paper is organized as follows. Section~2 reviews related work and motivation; Section~3 introduces AAG; Section~4 presents an initial validation; Section~5 discusses future directions; and Section~6 concludes.

%% file: sec2.tex
\begin{figure*}[t]
  \centering
  \includegraphics[width=0.9
\linewidth]{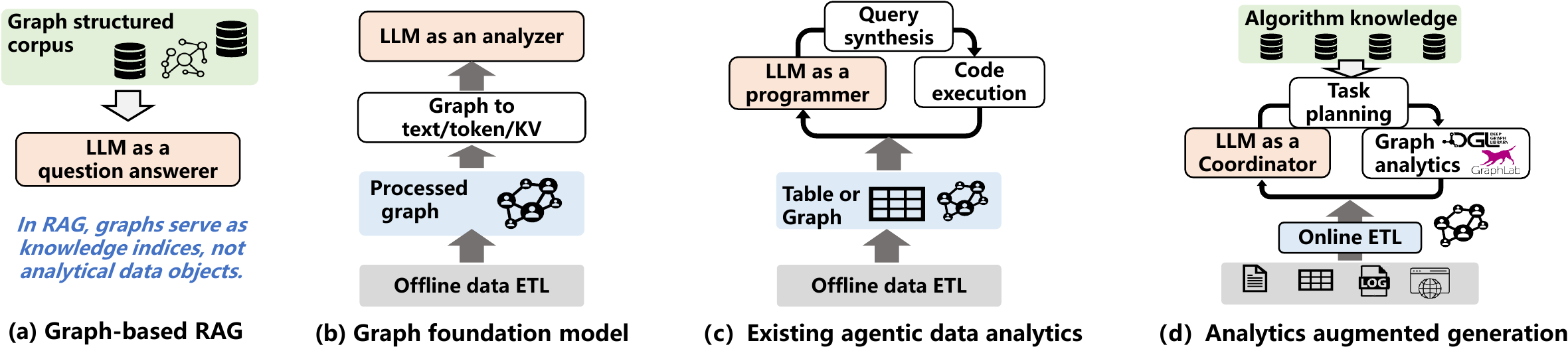}
  \caption{Comparison of Analytics-Augmented Generation (AAG) with existing approaches. 
  }
  \label{fig:sec1_comparison}
\end{figure*}

    

    

\section{Background and Motivations}

\subsection{Related Work}


\Paragraph{Graph-based RAG} Graph-based RAG~\cite{MGraphRAG_arxiv_2024,GRetriever_neurips_2024,ToG_iclr_2024,LightRAG_arxiv_2024,HippoRAG_neurips_2024,ragflow} incorporate graph structures primarily as indexing mechanisms for knowledge retrieval, focusing on question answering rather than data-centric graph analysis, as shown in Figure \ref{fig:sec1_comparison}(a).

\Paragraph{Graph Foundation Models (GFMs)}
Due to the limitations of the attention mechanism, LLM-centric analysis alone struggles to capture the semantics of graph-structured data. GFMs~\cite{GFMSurvey_arxiv_2025, GraphGPT_SIGIR_2024, GPT4Graph_arxiv_2023,GraphWiz_kdd_2024,GraphTMI_naaclhlt_2024,GraphInstruct_arxiv_2024,LLM4Graph_neurips_2024, GraphAgent_emnlp_2025} address this issue by introducing encoding/ graph2text/ graph2token modules that transform graphs into key–value or textual representations. However, such approaches rely on specialized encoders and exhibit limited generalization beyond predefined settings. Moreover, their scalability (input graph size) is fundamentally constrained by the limited context window of LLMs, as shown in Figure~\ref{fig:sec1_comparison}(b).

\Paragraph{Agentic data analytics over structural data}
A substantial body of LLM-based data agents focuses on natural language–to–SQL analytics over relational databases \cite{jiang2025siriusbi,pourreza2023din,li2025deepeye,DataAgentSurvey_ARXIV_2025}. Beyond SQL synthesis, recent data-agent frameworks have further explored natural language–driven, end-to-end analytics over structured data, often through multi-agent coordination, iterative refinement, and code-based execution \cite{liu2025joyagent,hong2025data,guo2024ds,zhang2025deepanalyzeagenticlargelanguage,sun2025agenticdata}. Representative systems include AgenticData \cite{sun2025agenticdata}, which emphasizes LLM-driven task planning and multi-agent collaboration for heterogeneous data analytics, and DeepAnalyze \cite{zhang2025deepanalyzeagenticlargelanguage}, which further investigates autonomous pipeline execution through agentic training and data-grounded reasoning. SiriusBI \cite{jiang2025siriusbi} integrates multiple fixed analytical modules with multi-round SQL generation to support reliable, industrial-scale business intelligence. Despite their effectiveness, these approaches primarily rely on the internal knowledge of (fine-tuned) LLMs for analytical decision making, as shown in Figure~\ref{fig:sec1_comparison}(c).

\Paragraph{Agentic data analytics over graph data}
Recent studies have explored agent-based methods for graph data analytics. \cite{GraphAgentReasoner_arxiv_2024, GraphTeam_arxiv_2024, chat2graph}. For example, GraphAgent-Reasoner \cite{GraphAgentReasoner_arxiv_2024} distributes graph data across multiple agents, where each agent manages a graph node and reasons collaboratively with its neighbors, enabling scaling to larger graphs. GraphTeam \cite{GraphTeam_arxiv_2024} organizes multiple specialized agents to handle sub-tasks such as query understanding, external knowledge retrieval, and problem-solving via code generation or direct reasoning. Chat2Graph \cite{chat2graph} combines the strengths of graph databases and agents to enable end-to-end execution from natural language to graph querying and analysis.

\begin{table}[t]
  \footnotesize
  \caption{Comparison of our proposal with representative work on agentic graph data analytics.}
  \vspace{-0.08in}
  \label{tab:commands}

  \renewcommand{\arraystretch}{0.9}
  \setlength{\tabcolsep}{3pt}

  \begin{adjustbox}{max width=\columnwidth, max height=0.45\textheight, keepaspectratio}
  \begin{tabular}{c|c|c|c|c}
    \hline

    \hline
    & \textbf{GraphAgent-} & \textbf{GraphTeam} & \textbf{Chat2Graph} & \multirow{2}*{\textbf{Our Proposal}} \\
    & \textbf{Reasoner}~\cite{GraphAgentReasoner_arxiv_2024} & \cite{GraphTeam_arxiv_2024}  & \cite{chat2graph}  & \\
    \hline
    Problem solving &
      \cellcolor{gray!25} Imperative &
      \cellcolor{gray!25} Imperative &
      \cellcolor{gray!25} Imperative &
      \cellcolor{gray!70} Declarative \\

    Input data &
      \cellcolor{gray!25} Pre-processed &
      \cellcolor{gray!25} Pre-processed &
      \cellcolor{gray!70} Raw &
      \cellcolor{gray!70} Raw \\

    Analysis capability &
      \cellcolor{gray!25} GAS-like API &
      \cellcolor{gray!70} Algorithm &
      \cellcolor{gray!25} GQL &
      \cellcolor{gray!70} Algorithm \\

    Multi-step support &
      \cellcolor{gray!25} Low &
      \cellcolor{gray!25} Low &
      \cellcolor{gray!25} Low &
      \cellcolor{gray!70} High \\

    Result scalability &
      \cellcolor{gray!25} Low &
      \cellcolor{gray!25} Low &
      \cellcolor{gray!25} Low &
      \cellcolor{gray!70} High \\
    \hline

    \hline
  \end{tabular}
  \end{adjustbox}
\end{table}

\subsection{Motivations}

Autonomous graph analytics has the potential to fundamentally broaden the accessibility and impact of graph-based data analysis. In practice, relational and tabular analytics have long dominated data-driven applications \cite{Andy_report_2024}, while rich relational signals embedded in large-scale data are often overlooked or underutilized due to the high cost of graph construction and analysis \cite{seattle_CACM_2022}. Advancing autonomous graph analytics can significantly lower the barrier of mining deep relational value hidden in data, and create substantial opportunities for real-world applications and emerging markets.

Despite this promise, existing solutions remain far from supporting practical end-to-end graph data analytics. As summarized in Table~\ref{tab:commands}, representative designs exhibit a set of systematic limitations that cannot be attributed to isolated engineering choices. Instead, these limitations stem from fundamental mismatches between the intrinsic structure of graph analytics and the ways current agent frameworks reason about analytical tasks and interact with underlying analytics systems.

\Paragraph{Lack of declarative problem solving}
As shown in Table \ref{tab:commands}, most existing graph agents adopt an imperative interaction paradigm, requiring users to explicitly specify concrete graph operations or algorithms (e.g., Hamiltonian cycle, PageRank, or GNN training). This design exposes low-level analytical details to users and assumes expertise in graph algorithms. In practice, however, users typically express declarative intent, what they want to analyze, rather than how to execute the analysis. The lack of declarative problem solving therefore, places the burden of intent-to-algorithm translation on users or on the LLM’s internal knowledge, which is brittle and difficult to extend.

\Paragraph{Rigid input requirement}
Table \ref{tab:commands} further highlights that most existing systems assume pre-constructed graphs as input. This assumption obscures a critical challenge in real-world analytics: graph data rarely exists in a ready-to-use form and must be extracted from heterogeneous raw sources. Moreover, appropriate graph schemas and representations are highly task-dependent. Requiring users to manually construct graphs not only increases engineering overhead but also tightly couples analytical logic with data modeling decisions, limiting reuse across tasks.

\Paragraph{Limited analysis capability}
Graph analytics lack a unified abstraction that supports general-purpose scheduling and execution for diverse algorithms \cite{GraphScope_VLDB_2021}. Consequently, existing agents either rely on restricted programming models (e.g., vertex-centric GAS\cite{GraphAgentReasoner_arxiv_2024} or GQL \cite{chat2graph}) or online code generation against fixed APIs such as NetworkX \cite{GraphTeam_arxiv_2024}, as shown in Table \ref{tab:commands}. While these abstractions simplify execution, they lack the expressiveness needed to compose heterogeneous graph algorithms into complex analytical workflows. More critically, online code generation becomes increasingly fragile in multi-stage graph analysis, where errors compound across steps: even with a per-step accuracy of 90\%, a four-step workflow incurs an overall failure rate of approximately 34\%, making multi-step analytics unreliable and difficult to validate.

\Paragraph{Multi-step workflow support}
Realistic graph analytics often involve multiple stages with explicit data dependencies, where the output of one step (e.g., a suspicious subgraph) becomes the input to subsequent analysis or learning tasks \cite{ByteGraph_VLDB_2022}. However, most existing systems treat analytical steps in isolation \cite{GraphAgentReasoner_arxiv_2024,GraphTeam_arxiv_2024} or provide limited support for workflow-level planning and coordination (Chat2Graph primarily supports multi-step retrieval and query workflows). As a result, even when individual algorithms are supported, composing them into coherent, dependent workflows remains challenging.

\Paragraph{Limited Analysis Scalability}
Finally, graph analytics produce large and semantically rich outputs, including subgraphs, rankings, and node- or edge-level predictions. As summarized in Table \ref{tab:commands}, existing agents typically return raw or lightly processed results to the LLM, which quickly exceeds context budgets and hinders effective reasoning for large-scale data. This limitation fundamentally constrains scalability and prevents LLMs from participating meaningfully in multi-stage analytical reasoning.


\Paragraph{Summary}
Taken together, these limitations reveal a fundamental design gap: existing graph agents implicitly assume the users with sufficient domain knowledge to manually perform certain key steps of graph analysis. They lack explicit analytical grounding for translating intent into execution, failing to leverage the rich body of graph algorithm knowledge, to support verifiable execution across diverse graph algorithms, or to enable task-aware automated data preparation. This observation motivates a shift toward a new design paradigm that bridges declarative user intent and reliable large-scale graph analytics. We propose Analytics-Augmented Generation (AAG), which elevates analytical computation to a first-class concern and positions LLMs as knowledge-grounded analytical coordinators. As illustrated in Figure~\ref{fig:sec1_comparison}(d), AAG enables LLMs to interpret high-level user intent, synthesize executable analytical workflows, coordinate task-aware graph construction, and orchestrate algorithm execution under explicit analytical knowledge, thereby supporting natural language-driven end-to-end graph analytics.

%% file: sec3.tex
\begin{figure}
  \centering
  \includegraphics[width=\linewidth]{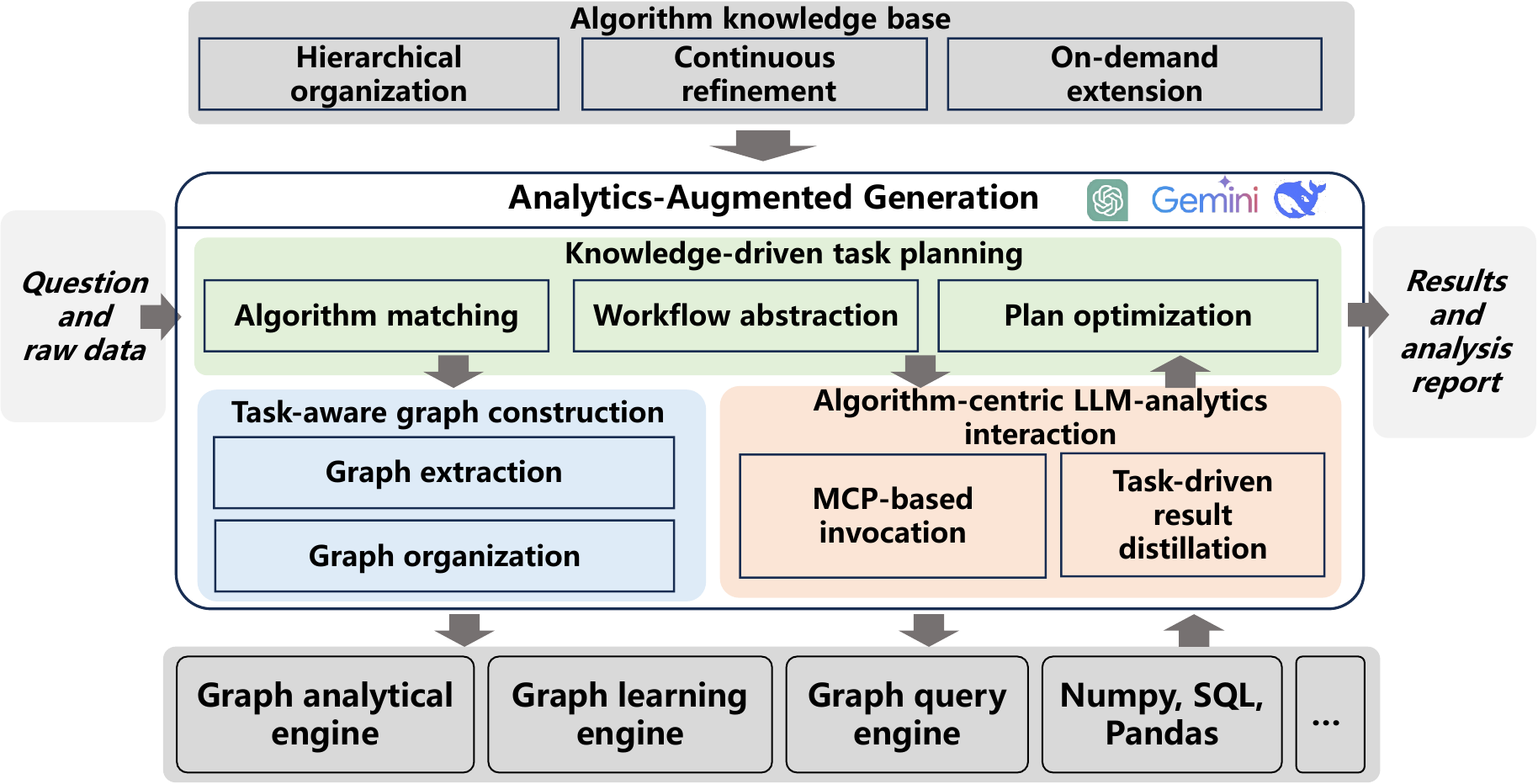}
  \vspace{-0.1in}
  \caption{Analytics-augmented generation overview. 
  }
  \label{fig:sec3_system}
  \vspace{-0.1in}
\end{figure}

\section{Analytics-Augmented Generation }

AAG is organized around three decoupled components and an external knowledge base, as shown in Figure~\ref{fig:sec3_system}.
The \textbf{knowledge-driven task planning} (Sec.~3.2) layer grounds declarative natural-language requests in a purpose-built and extensible \textbf{hierarchical knowledge base} (Sec.~3.1), enabling automatic translation of high-level user intent into multi-stage analytical workflows.
The \textbf{algorithm-centric LLM–analytics interaction} (Sec.~3.3) layer executes workflow stages via standardized invocation protocols over mature graph systems, while transforming execution results into LLM-consumable representations.
The \textbf{task-driven graph construction} (Sec.~3.4) layer models and extracts graph data based on task semantics and inter-stage dependencies, dynamically preparing task-relevant data structures during execution.

%% file: sec4.tex
\subsection{Hierarchical algorithm knowledge base}
\begin{figure}
  \centering
  \includegraphics[width=0.9\linewidth]{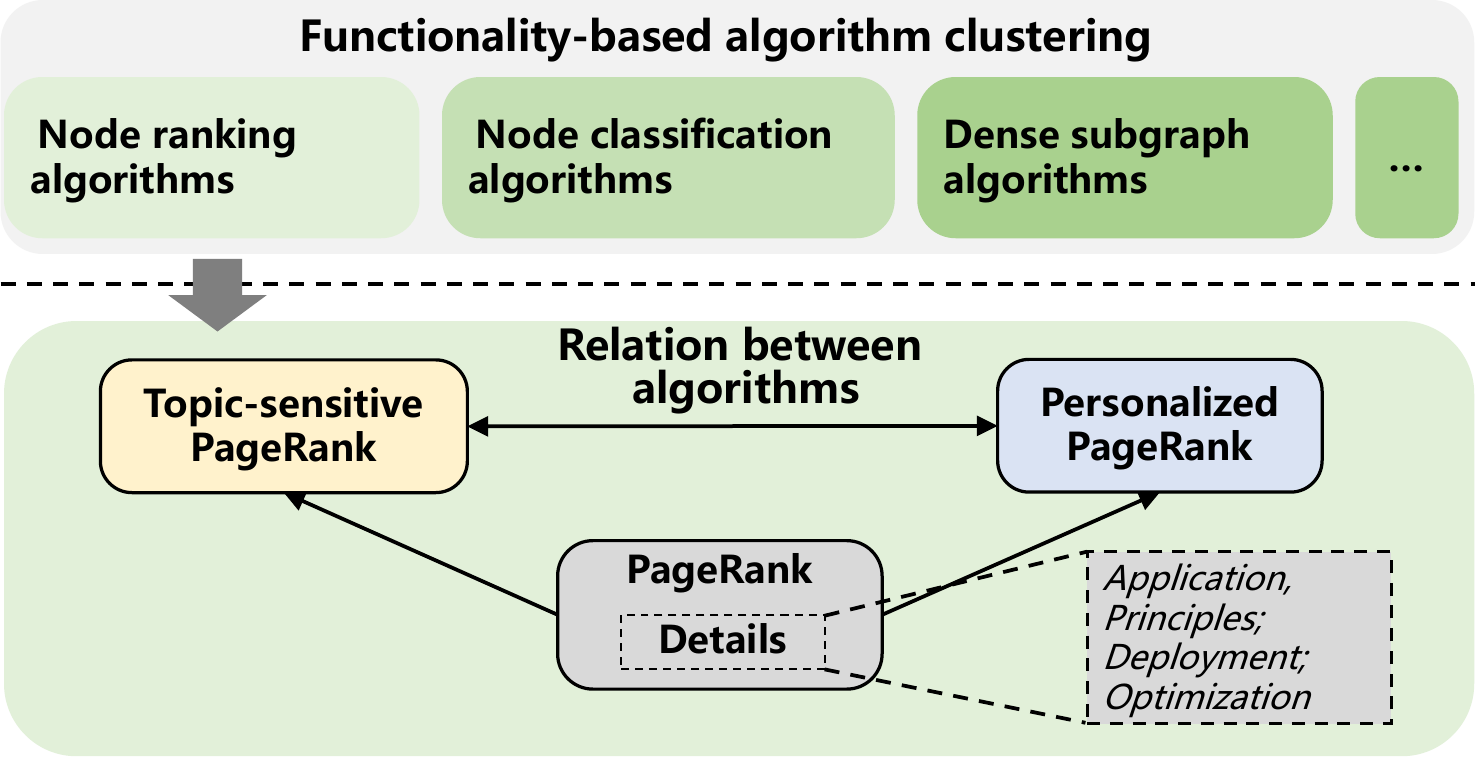}
  \caption{Hierarchical knowledge organization for efficient retrieval of massive graph algorithms. }
  \label{fig:sec3_knowledge}
\end{figure}

The knowledge base provides semantic and algorithmic grounding for planning and orchestrating graph analytics tasks.  It organizes analytical knowledge in a structured and queryable form that supports reasoning over algorithm choices, applicability, and trade-offs. The design is centered on representing graph analytics knowledge in a hierarchical, graph-structured manner, enabling coarse-to-fine retrieval during task planning. Figure \ref{fig:sec3_knowledge} shows the overview.

\Paragraph{Hierarchical knowledge graph} 
To enable scalable and intent-aware algorithm discovery, the graph analytics knowledge should be organized around a hierarchical abstraction rather than flat collections. The algorithms should be grouped into functional families that reflect shared analytical objectives and usage patterns, allowing task planning to reason at the level of algorithm classes before committing to specific methods. Structural relations within each algorithm family should capture meaningful distinctions between algorithms ,such as personalization or topic sensitivity in ranking algorithms, enabling the planner to navigate the space through structured traversal instead of exhaustive retrieval. Detailed algorithm information is associated with leaf nodes and accessed only after high-level candidates are selected, exposing low-level details in early planning stages.

\Paragraph{Continuous knowledge refinement}
A practical challenge in constructing high-quality algorithmic knowledge from papers and system documentation is the presence of substantial non-essential content, which degrades retrieval quality and increases planning overhead. Purely rule-based filtering or unguided LLM-based pruning may either miss relevant information or remove useful content. A promising direction is to incorporate user feedback into knowledge refinement by allowing users to inspect retrieved knowledge fragments during task planning and assess their contribution to the analytical plan manually or automatically. These signals are used to estimate the relative usefulness of different knowledge regions and guide future retrieval and filtering, progressively deprioritizing low-value content and focusing planning on high-value knowledge.

\Paragraph{Dynamic knowledge base expansion}
Given the rapid evolution of graph analytics, maintaining a complete and up-to-date knowledge base that covers all algorithm progress is neither feasible nor desirable. Instead, the AAG emphasizes on-demand knowledge expansion: when existing knowledge is insufficient, the planning layer should trigger a web-search module to retrieve relevant literature using keyword queries and citation links. Candidate algorithms are then selectively incorporated into the knowledge base. By restricting expansion to task-relevant methods while preserving the hierarchical organization, this approach enables continuous adaptation without the cost of indiscriminate knowledge ingestion.

\begin{figure}
  \centering
  \includegraphics[width=\linewidth]{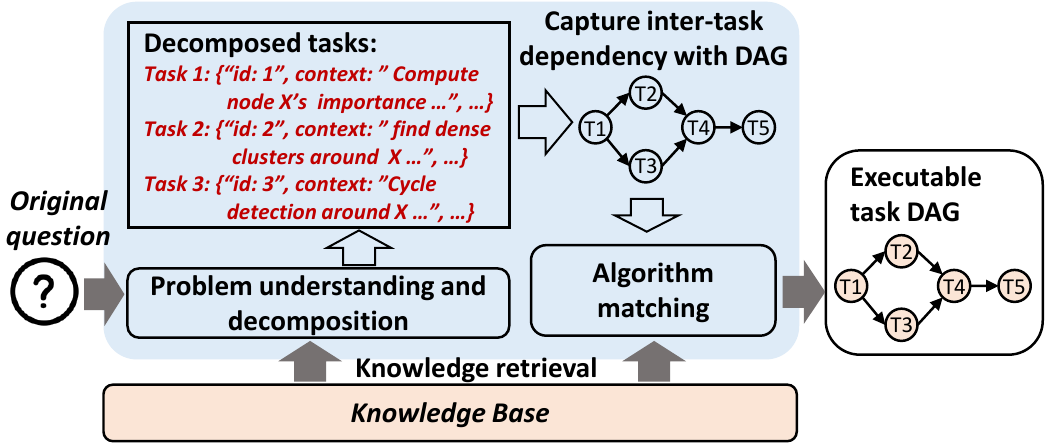}
  \vspace{-0.1in}
  \caption{The original problem is transformed into an executable task DAG using the external knowledge base.} 
  \label{fig:sec4_decision}
  \vspace{-0.1in}
\end{figure}

\subsection{Knowledge-Driven Task-planning}
The task planning layer bridges open-ended user intent and executable graph analytics workflows. Given a natural-language query describing a complex problem, the planning process incrementally interprets the intent, decomposes the task, identifies relevant classes of graph analytics algorithms, and organizes them into a structured workflow for execution, as illustrated in Figure~\ref{fig:sec4_decision}.

\Paragraph{Algorithm matching}
Mapping natural-language intent to graph analytics decisions requires both semantic understanding and structural reasoning over analytical knowledge. The task planning layer therefore combines semantic retrieval with structured traversal over the algorithm knowledge organization: semantic similarity identifies user-requirement relevant algorithm families, while structural relations support reasoning about applicability and compatibility. Fine-grained algorithm attributes are then consulted to refine candidate choices, such as constraints, assumptions, and configuration considerations. This integrated design can effectively reduce the reliance on purely model-internal reasoning and enables more grounded analytical planning.

\Paragraph{Workflow abstraction}
Once candidate analytical components are identified, the planning layer organizes them into an explicit workflow represented as a Directed Acyclic Graph (task DAG). Graph nodes correspond to modular analytical algorithms, while edges encode data dependencies and execution dependency (the input data, parameter, and decision comes from the output of the previous algorithm). This representation makes analytical workflows explicit and interpretable, and decouples high-level analytical reasoning from low-level execution details, allowing workflows to be validated, adapted, or partially reconfigured without redefining the entire plan.

\Paragraph{Execution plan refinement}
Although the pre-generated task graph is constructed based on carefully retrieved knowledge, complex graph analytics tasks may still encounter suboptimal parameter choices, execution failures, or incompatible algorithm combinations, issues that often emerge only during actual execution. To accommodate such uncertainty, a feedback-guided optimization loop can be incorporated. Execution outcomes and result quality provide feedback signals to revise workflows, adjust configurations, or explore alternative algorithms, allowing workflows to progressively converge toward more effective solutions.

\begin{figure}
  \centering
  \includegraphics[width=0.8
  \linewidth]{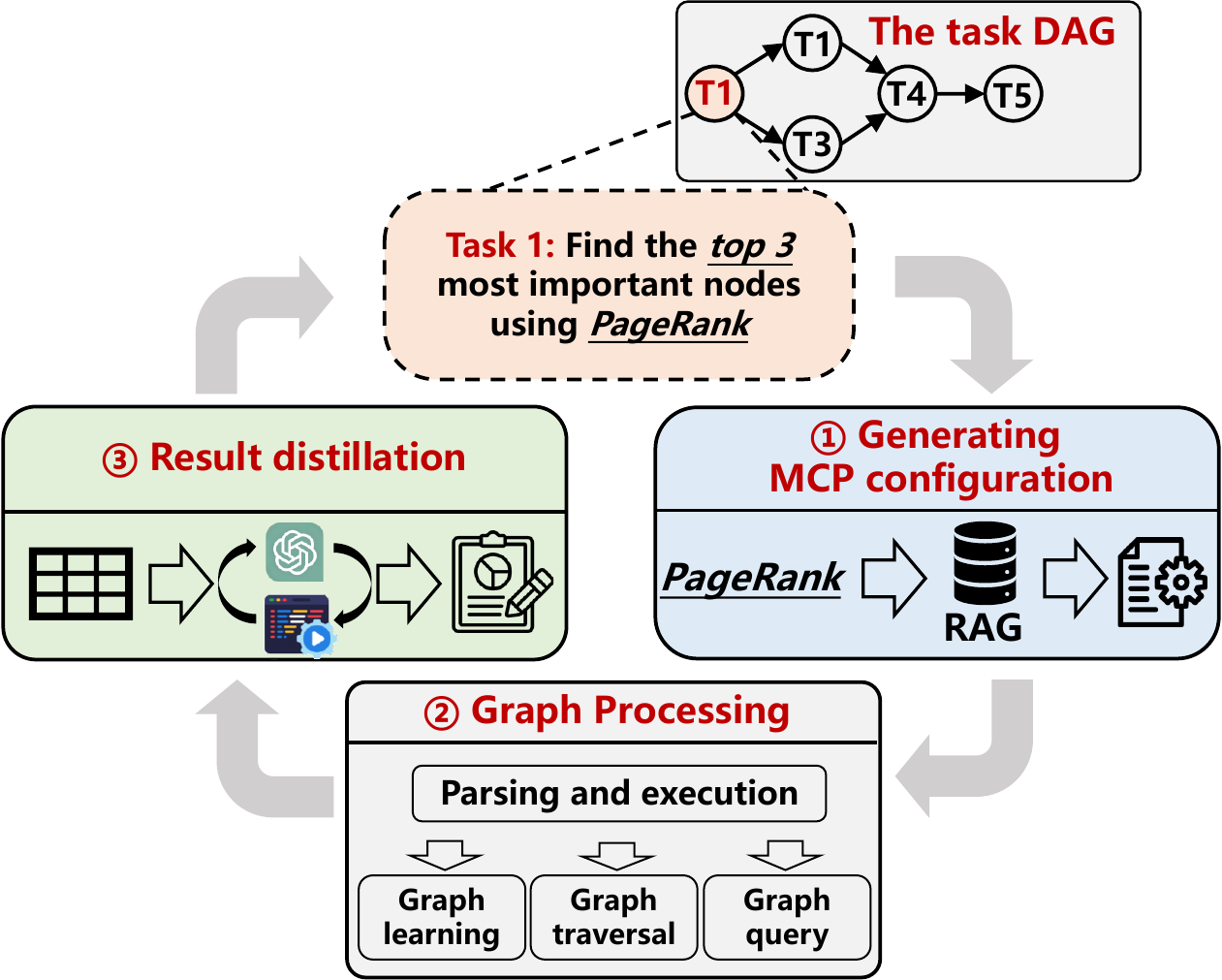}
  \vspace{-0.1in}
  \caption{The algorithm-centric interaction between the LLM and Graph engine.
  }
  \label{fig:sec4_interaction}
\end{figure}

\subsection{Algorithm-centric LLM-analytics Interaction}

The interaction layer bridges high-level analytical reasoning and concrete graph execution engines. As illustrated in Figure~\ref{fig:sec4_interaction}, it operates over the abstract task DAG, translating each analytical step into executable graph operations. For each task node, the LLM configures and invokes the corresponding algorithm on appropriate graph engines, then filters execution outputs and returns only task-relevant results to the workflow.

\Paragraph{MCP-based Algorithm-centric invocation}
A key challenge in LLM-driven graph analytics is enabling automated and reliable invocation of diverse graph algorithms without exposing low-level system details. The interaction layer therefore adopts an Model Context Protocol\footnote{https://modelcontextprotocol.io/docs/getting-started/intro} (MCP)-based algorithm-centric invocation abstraction, in which algorithms configurations are described through structured interfaces capturing functionality, required inputs, and execution constraints. Under this abstraction, the LLM selects and parameterizes analytical operations through RAG, while execution details are handled by underlying graph engines, keeping analytical reasoning decoupled from system-specific implementations.

\Paragraph{Task-driven result distillation}
Graph algorithms often produce large and semantically heterogeneous outputs (e.g., node rankings, classification outputs, or matched subgraphs) that cannot be directly consumed by LLMs due to context limits. The interaction layer therefore requires a semantics-driven result mediation module that distills raw outputs into concise, task-relevant summaries by filtering irrelevant data and extracting salient insights (e.g., top-K PageRank values, cyclical structures). By exposing only distilled evidence, this design supports effective multi-stage LLM–analytics interaction under limited context budgets.

\subsection{Task-driven Graph Construction}
The graph construction layer bridges  high-level task semantics and algorithmic requirements with execution-ready graph representations. Rather than a one-time preprocessing step, graph construction is treated as a task-driven process shaped by analytical intent and algorithm characteristics. For each task, the layer selects relevant schema elements and organizes graphs to support efficient computation, enabling adaptation across analytical tasks while keeping data preparation transparent to users.

\Paragraph{Task-aware graph extraction}
Graph data rarely exist in an analysis-ready form and must be constructed from heterogeneous sources based on task requirements. The graph construction layer therefore emphasizes task-aware schema derivation: given a user query, the system infers relevant entities, relations, and attributes and aligns them with algorithmic knowledge to derive a task-specific graph schema. This design is based on the observation that a single data source (massive transaction records) may often supports multiple analytical tasks with different semantic needs, for example, fraud detection focuses on user-to-user money flows, while recommendation relies on user-to-merchant purchase relations. By restricting graph construction to task-relevant data, schema-focused extraction avoids irrelevant semantics interference across tasks and reduces downstream processing cost, providing a concise specification for subsequent graph construction.

\Paragraph{Execution-aware graph construction} Graph analytics workflows pose two fundamental challenges for graph construction. First, different graph algorithms impose distinct data layout and access requirements, making a single fixed representation inefficient across workflow stages. Second, analytical workflows exhibit inter-stage data dependencies, where downstream inputs may depend on intermediate results from upstream task rather than solely on the original data. For example, in fraud detection, a GNN may operate on a suspicious subgraph identified by an upstream stage instead of the full transaction graph. To address these challenges, graph construction should be treated as a task-aware and stage-aware process that adapts graph organization to algorithm execution characteristics and inter-stage data dependencies. This approach avoids the inefficiencies introduced by relying on a single fixed representation, thereby enabling more flexible and efficient analytical execution.

%% file: sec5.tex
\begin{figure*}
  \centering
  \includegraphics[width=\textwidth]{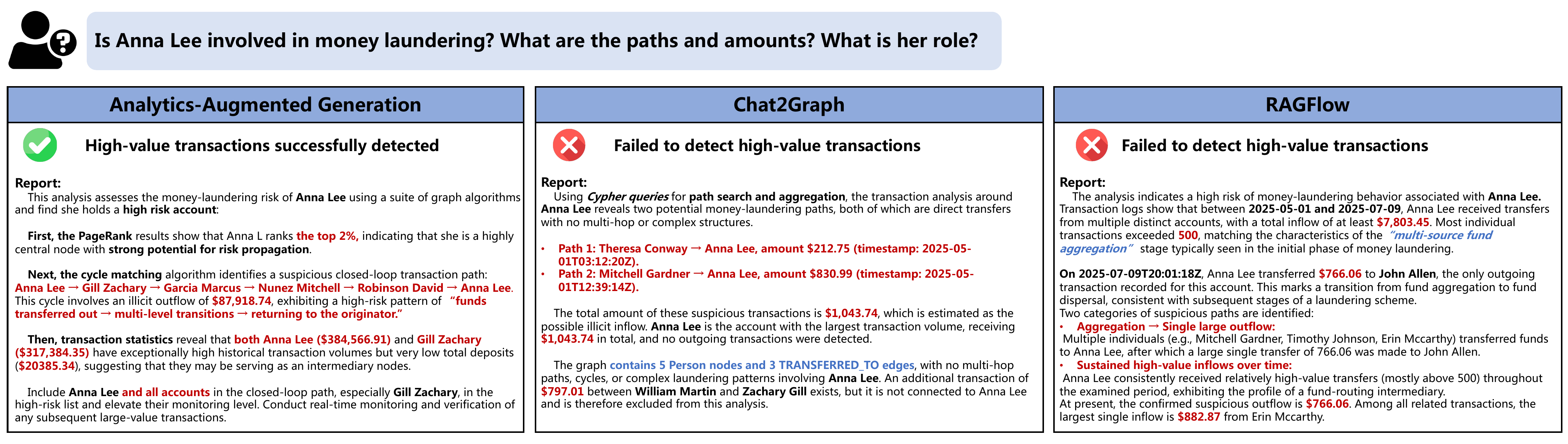}
  \caption{Comparison of the AAG against Chat2Graph \cite{chat2graph} and RAGFlows \cite{ragflow}.} 
  \label{fig:sec5_compare}
\end{figure*}

\section{Prototype and Case Study }


We implement a lightweight prototype to provide an initial validation of the AAG paradigm. The prototype is built on widely used graph analytics systems, including DGL for graph learning, NetworkX for graph querying, and GraphScope for analytical processing, and leverages their built-in algorithm implementations as execution backends. In addition, NumPy and Pandas are integrated for tabular processing and result extraction. This prototype is not intended as a full system realization, but as a proof of concept demonstrating the feasibility of AAG-style coordination over existing graph analytics engines. 

GPT-4o mini is employed as the underlying model for coordination. The knowledge base is constructed from system documentation (e.g., README files) and corresponding research papers, following the knowledge organization described in Section 4.2. Its dynamic maintenance and refinement are left as future work. The current prototype implements an initial MCP-based algorithm invocation interface and task-driven result extraction mechanisms, and supports task-aware graph construction for a set of commonly used graph algorithms. Workflow construction currently follows a fixed approach, while adaptive optimization based on reinforcement learning is deferred to future work.

We evaluate the prototype using the money laundering detection scenario illustrated in Figure~\ref{fig:sec1_problem}. Experiments are conducted on an open financial transaction dataset (the IBM AMLSim example dataset \cite{AMLSim_ibm_2021}), which contains 1,446 users and 17,512 transactions with simulated high-value transactions. We compare AAG with a representative graph agent, Chat2Graph \cite{chat2graph}, and a graph-based retrieval-augmented generation (RAG) system, RAGFlow \cite{ragflow}.

Compared with existing approaches, a key advantage of AAG lies in its ability to interpret user intent and decompose it into a coherent multi-stage analytical workflow. In the money laundering detection task, AAG decomposes the analysis into four stages: (1) assessing whether Anna Lee is a high-risk user; (2) identifying potential money-laundering cycles involving Anna Lee; (3) estimating the amount of illicit fund transfers along these cycles; and (4) summarizing Anna Lee’s incoming and outgoing transactions. Based on this decomposition, AAG selects and orchestrates appropriate analytical components, including graph algorithms such as PageRank and cycle detection, as well as NumPy-based aggregation for transaction summarization. In contrast, existing systems lack algorithmic knowledge grounding and intent-level understanding, and therefore resort to shallow analytical strategies, such as simple two-hop neighborhood queries. As shown in Figure~\ref{fig:sec5_compare}, Chat2Graph fails to construct an appropriate analytical workflow and only analyzes a limited subset of transactions around Anna Lee. RAGFlow, which focuses on single-pass retrieval, similarly cannot capture the multi-stage analytical structure required by the task. Consequently, neither system can identify high-value transaction paths, whereas AAG successfully discovers relevant high-value transaction cycles and provides step-by-step analytical evidence to support interpretability.

\section{Future Development of AAG}
\Paragraph{Benchmarking end-to-end analytics} Existing benchmarks for LLM-assisted graph processing mainly focus on single-hop tasks~\cite{chatgraph_icde_2025,LLM4Graph_NIPS_2024}, which fail to reflect the complexity of real-world end-to-end applications. In practice, graph analytics often involve multi-stage workflows that combine graph computation with tabular result analysis, motivating benchmarks that evaluate planning quality, execution correctness, and end-to-end performance.

\Paragraph{Shared and federated analytical knowledge}
Across users, substantial portions of domain knowledge (e.g., financial graph analytics)~\cite{dupin_sigmod_2025} are inherently shared. Reconstructing similar knowledge bases independently incurs unnecessary token and computational overhead. A trustworthy federated knowledge-management layer~\cite{Fedgraphdata_book_2011} can enable sharing and reuse of knowledge subgraphs, reducing redundancy while improving efficiency.

\Paragraph{Community-driven algorithm module maintenance}
AAG currently assumes a sufficiently rich set of execution engines, with new engines manually integrated when needed. To reduce this burden, a community-maintained application pool~\cite{crowdsource_SIGMOD_2017} can support automated integration of new algorithm modules, improving coverage and usability.

\Paragraph{Efficient execution for AAG}
AAG integrates LLM inference, graph computation, retrieval, and graph learning into dynamic and heterogeneous pipelines, imposing higher scheduling and resource-management demands than traditional LLM+RAG serving~\cite{RAG1_SOSP_2025,RAG2_SOSP_2025}. Resource-disaggregated architectures~\cite{LLMDisaggregate_VLDB_2024} and fine-grained task scheduling frameworks~\cite{ByteScheduler_SOSP_2019} offer promising directions for improving execution efficiency.

\Paragraph{Extension to other fields} 
In addition, we emphasize that the AAG architecture is not limited to graph data analytics, but can be naturally extended to other scientific domains by integrating large bodies of domain literature and specialized analytical tools, advancing autonomous problem solving for complex tasks in natural sciences and engineering.

\section{Conclusion}

We argue that practical autonomous graph data analytics cannot be realized by retrieval- or code-centric LLM agents alone. The fundamental challenge lies in the absence of an end-to-end abstraction that coherently connects high-level analytical intent, task-aware graph construction, and reliable execution across diverse graph algorithms. In this work, we envision Analytics-Augmented Generation (AAG) as a step toward intent-driven graph analytics, grounded in external knowledge–based planning, algorithm-centric interaction, and task-aware graph construction. Rather than replacing existing graph systems, AAG builds upon their reliability while elevating LLMs to the role of decision-making coordinators. We believe AAG outlines a promising research direction for graph data agents and the co-design of reasoning, analytics, and graph data management systems.